\documentclass[fleqn,10pt]{wlscirep}
\usepackage[utf8]{inputenc}
\usepackage[T1]{fontenc}

\title{Comparing Alternatives to the Fixed Degree Sequence Model for Extracting the Backbone of Bipartite Projections}

\author[1,*]{Zachary P. Neal}
\author[2]{Rachel Domagalski}
\author[2]{Bruce Sagan}
\affil[1]{Michigan State University, Psychology Department, East Lansing MI, USA}
\affil[2]{Michigan State University, Mathematics Department, East Lansing MI, USA}
\affil[*]{zpneal@msu.edu}

\usepackage{amsmath,amssymb,amsfonts,amsthm,array,tikz,courier,soul}

\newtheorem{thm}{Theorem}[section]

\theoremstyle{definition}

\newcommand{\bprf}{\begin{proof}}
\newcommand{\eprf}{\end{proof}\medskip}
\newcommand{\dil}{\displaystyle}
\newcommand{\ree}[1]{(\ref{#1})}
\newcommand{\al}{\alpha}


\begin{abstract}
Projections of bipartite or two-mode networks capture co-occurrences, and are used in diverse fields (e.g., ecology, economics, bibliometrics, politics) to represent unipartite networks. A key challenge in analyzing such networks is determining whether an observed number of co-occurrences between two nodes is significant, and therefore whether an edge exists between them. One approach, the fixed degree sequence model (FDSM), evaluates the significance of an edge's weight by comparison to a null model in which the degree sequences of the original bipartite network are fixed. Although the FDSM is an intuitive null model, it is computationally expensive because it requires Monte Carlo simulation to estimate each edge's $p$-value, and therefore is impractical for large projections. In this paper, we explore four potential alternatives to FDSM: fixed fill model (FFM), fixed row model (FRM), fixed column model (FCM), and stochastic degree sequence model (SDSM). We compare these models to FDSM in terms of accuracy, speed, statistical power, similarity, and ability to recover known communities. We find that the computationally-fast SDSM offers a statistically conservative but close approximation of the computationally-impractical FDSM under a wide range of conditions, and that it correctly recovers a known community structure even when the signal is weak. Therefore, although each backbone model may have particular applications, we recommend SDSM for extracting the backbone of bipartite projections when FDSM is impractical.
\end{abstract}
\begin{document}

\flushbottom
\maketitle
\thispagestyle{empty}

\section*{Introduction}
Bipartite or two-mode networks are composed of two types of nodes, which we call \textit{agents} and \textit{artifacts}, and edges between nodes of one type and nodes of the other type. These networks can be used to represent a wide range of phenomena and therefore are studied in a diverse range of disciplines. For example, natural selection unfolds as species (the agents) compete over sites (the artifacts), commerce is possible as traders exchange resources, scientific advances are reported as scholars write papers, and laws are adopted as legislators sponsor bills. Although bipartite networks are useful in their own right, they can also be useful for inferring unipartite (i.e., one-mode) networks that are difficult to measure directly. For example, while it may be difficult to directly survey politicians about their political alliances because they are busy and may have reasons to misrepresent their true alliances, it may be possible to infer political alliances from politicians' co-sponsorship of legislation, which is readily observable.\cite{neal2020,fowler2006legislative} A bipartite projection transforms a bipartite network into a unipartite co-occurrence network in which pairs of agents are connected by edges whose weights capture their number of shared artifacts.\cite{saracco2015randomizing,diclemente2021urbanization,simmons2019bmotif} For example, competitive interaction networks can be inferred from species' co-occurrence in sites,\cite{diamond1975assembly} trade networks can be inferred from firm co-location\cite{taylor2002measurement,straka2017grand,saracco2017inferring} or product co-exchange,\cite{saracco2015randomizing} scholarly collaboration networks can be inferred from paper co-authorship,\cite{newman2001} and political alliance networks can be inferred from bill co-sponsorship.\cite{neal2020} Throughout the paper we use these applications to offer concrete examples, however the models we discuss are general and can be applied to derive unipartite backbones in such diverse contexts as flavor,\cite{ahn2011flavor} misinformation,\cite{tollefson2021tracking} text,\cite{radhakrishnan2017novel} and genetic\cite{zhang2005} networks. Indeed, in principle any unipartite network can be represented as the projection of some bipartite network.\cite{filho2020,guillaume2004bipartite,newman2003social} 

Despite their promise, bipartite projections (i.e., co-occurrence networks) are challenging to analyse because they are typically dense and weighted, and because the edge weights do not necessarily capture the strength of the relationship between nodes.\cite{neal2014backbone} As a result, it is often useful to analyze the \textit{backbone} of a bipartite projection, which is an unweighted and typically sparser network that retains only the most `important' edges. Although well-known methods exist for extracting the backbone of weighted networks that are not bipartite projections,\cite{serrano2009extracting,dianati2016unwinding} methods designed specifically for bipartite projections have recently been developed.\cite{neal2014backbone,zweig2011systematic,saracco2017inferring,Tumminello} Among these methods, the fixed degree sequence model (FDSM) relies on an intuitive null model, but requires computationally expensive Monte Carlo simulations, making it impractical for extracting the backbone of large bipartite projections. Faster methods are available, however relatively little is known about whether they yield backbones that are similar to those that would be obtained from using FDSM,\cite{cimini2021meta} and therefore whether they offer computationally efficient alternatives. To offer guidance to researchers wishing to extract an FDSM-like backbone from a large bipartite projection, in this paper we consider four potential alternatives to FDSM: fixed fill model (FFM) fixed row model (FRM), fixed column model (FCM), and stochastic degree sequence model (SDSM).

The paper is organized in six sections. We begin by formally defining bipartite projections, backbones, and the five backbone models, presenting proofs of the probability mass functions for their respective edge weight distributions in the \textit{Supplementary Text S1}. In study 1, we evaluate the accuracy and speed of different approaches for estimating cell-filling probabilities used by the SDSM. 
In study 2, we evaluate the statistical power of the SDSM relative to the FDSM. In study 3, we examine how degree distributions impact the similarity of backbones extracted using FDSM and each of the alternative models. In study 4, we examine the extent to which backbones extracted using different models accurately recover a known community structure. Finally, we conclude with recommendations for backbone model selection and opportunities for future model development.

\section*{Backbone extraction for bipartite projections}
\subsection*{Preliminaries}
A \textit{bipartite network} captures connections between nodes of one type (\textit{agents}) and nodes of a second type (\textit{artifacts}). Throughout this section, we use the ecological case of Darwin's Finches to provide a concrete example.\cite{sanderson2000testing,gotelli2000null} On his voyage to the Galapagos Islands on the H.M.S. Beagle, Darwin observed that only some species of finches lived on each island. These patterns can be represented as a bipartite network in which finch species (the agent nodes) are connected to the islands (the artifact nodes) where they are found.\cite{neal_neal_2020} A bipartite network can be represented as a binary matrix in which the agents are arrayed as rows, and the artifacts are arrayed as columns. We use $\mathbf{B}$ to denote a bipartite network's representation as a matrix, where $B_{ik}=1$ if agent $i$ is connected to artifact $k$, and otherwise is $0$. The sequence of row sums and the sequence of column sums of $\mathbf{B}$ are called the agent and artifact degrees sequences, respectively. These sequences are among the bipartite network's most significant features and are known to have implications for bipartite projections and backbones.\cite{filho2020,domagalski2021backbone,neal2021} In the ecological case, the agent degree sequence captures the number of islands where each species is found, while the artifact degree sequence captures the number of species found on each island.

The \textit{projection} of a bipartite network is a weighted unipartite co-occurrence network in which a pair of agents is connected by an edge with a weight equal to their number of shared artifacts. For example, the bipartite projection of Darwin's finch network is a species co-occurrence network in which a pair of finch species is connected by an edge with a weight equal to the number of islands where they are both found. We use $\mathbf{P}$ to denote the matrix representation of a bipartite projection, which is computed as $\mathbf{BB}^T$, where $\mathbf{B}^T$ indicates the transpose of $\mathbf{B}$. In a projection $\mathbf{P}$, $P_{ij}$ indicates the number of times agents $i$ and $j$ were connected to the same artifact $k$ in $\mathbf{B}$. The diagonal entries of $\mathbf{P}$, $P_{ii}$, are equal to the agent degrees, but in practice are ignored.

The \textit{backbone} of a bipartite projection is a binary representation of $\mathbf{P}$ that contains only the most `important' or `significant' edges. For example, the backbone of a species co-occurrence network connects pairs of species if they are found on a significant number of the same islands, which might be interpreted as evidence that the two species do not compete for resources and perhaps are symbiotic. We use $\mathbf{P}'$ to denote the matrix representation of the backbone of $\mathbf{P}$. Because multiple methods exist for deciding when an edge is significant and thus should be preserved in the backbone, we use $\textbf{P}^{'\text{M}}$ denote a backbone extracted using method $M$. It is important to note that for a given bipartite projection, there is no `true' backbone, but only backbones corresponding to specific backbone methods $M$. The backbone extracted using FDSM (i.e. $\textbf{P}^{'\text{FDSM}}$) may be similar or different from a backbone extracted using another method such as SDSM (i.e. $\textbf{P}^{'\text{SDSM}}$), and these similarities and differences depend on the information that is considered by the respective methods when determining whether edges' weights are significant. It is these similarities and differences that we explore in the four studies below.

Backbone extraction methods that were originally developed for non-projection weighted networks are often also applied to weighted bipartite projections. One simple method preserves an edge in the backbone if its weight in the projection exceeds some \textit{global threshold} $T$. However, when $T = 0$, which is common, the backbone is very dense and has a high clustering coefficient because each artifact of degree $d$ induces $d(d-1)/2$ edges in the backbone.\cite{latapy2008} Using $T > 0$ can yield a sparser and less clustered backbone,\cite{derudder2005cliquishness,fong2020,bratton2011} but still yields highly clustered networks in which low-degree nodes are excluded while high-degree nodes are preserved.\cite{serrano2009extracting} More sophisticated methods, including the \textit{disparity filter}\cite{serrano2009extracting} and \textit{likelihood filter},\cite{dianati2016unwinding} aim to overcome these limitations of the global threshold method by using a different threshold for each edge based on a null model. However, all methods that can be applied to non-projection weighted networks have the same shortcoming when applied to weighted bipartite projections: they ignore information about the artifacts, which is lost when generating the projection. \cite{neal2014backbone} In the ecological case, the global threshold, disparity filter, and likelihood filter methods all decide whether two species should be connected in the backbone only by examining how many islands these two species are both found on, but do not consider the characteristics of those islands, including how many other species are found there, or even how many islands there are. Therefore, although these methods are promising for extracting the backbone from non-projection weighted networks, different methods are required for extracting the backbone from a bipartite projection.

\subsection*{Bipartite ensemble backbone models}
Bipartite ensemble backbone models decide whether an edge's observed weight $P_{ij}$ is significantly large, and thus whether a corresponding edge should be included in the backbone by comparing it to an ensemble of random bipartite networks. Let $\mathcal{B}$ be the set of all bipartite networks $\mathbf{B^*}$ having the same number of agents and artifacts as $\mathbf{B}$. In the ecological case, $\mathbf{B^*}$ might be viewed as representing a possible world containing the same species and islands, but in which locations of species on islands is different, and likewise $\mathcal{B}$ is the set of all such possible worlds. The bipartite ensembles used in backbone models take a subset $\mathcal{B^\text{M}}$ of $\mathcal{B}$, subject to certain constraints $M$, and impose a probability distribution on it. In all models except the SDSM, the uniform probability distribution is imposed on $\mathcal{B^\text{M}}$, that is, each element of the ensemble is equally likely. The backbone is then extracted from the projection of $\mathbf{B}$ by using the distribution of edge weights arising from projections of members of the ensemble to evaluate their statistical significance.

We use $P^*_{ij}$ to denote a random variable equal to $(\mathbf{B^*}\mathbf{B^*}^T)_{ij}$ for $\mathbf{B^*}~\in~\mathcal{B}^\text{M}$. That is, $P^*_{ij}$ is the number of artifacts shared by $i$ and $j$ in a bipartite network randomly drawn from $\mathcal{B}^\text{M}$. In the ecological case, $P^*_{ij}$ represents the number of islands that are home to both species $i$ and $j$ in a possible world, while the distribution of $P^*_{ij}$ is the distribution of the number of islands shared by species $i$ and $j$ in all possible worlds.

Decisions about which edges should appear in a backbone extracted at the statistical significance level $\alpha$ are made by comparing $P_{ij}$ to $P^*_{ij}$
\[ P_{ij}'=
\begin{cases}
1 & \text{ if } \Pr(P^*_{ij} \geq P_{ij}) < \frac{\alpha}{2},\\
0 & \text{otherwise.}
\end{cases}\]
This test includes edge $P'_{ij}$ in the backbone if its weight in the observed projection $P_{ij}$ is uncommonly large compared to its weight in projections of members of the ensemble $P^*_{ij}$. We use a two-tailed significance test in the studies below because, in principle, an edge's weight in the observed projection could be uncommonly \textit{larger} or uncommonly \textit{smaller} than its weight in projections of members of the ensemble, however a one-tailed test may also be used. In the ecological case, two species are connected in the backbone if their number of shared islands in the observed world is uncommonly large compared to their number of shared islands in all possible worlds. 

There are many ways that $\mathcal{B}$ can be constrained,\cite{strona2018bi} with each set of constraints describing a particular ensemble $\mathcal{B}^\text{M}$, which is used in a particular ensemble backbone model $M$ to yield a particular backbone $\mathbf{P}^{'M}$. In the case of ensembles used to extract the backbone of bipartite projections, our focus in this paper, two broad types of constraints are common.\cite{cimini2021meta} First, ensembles can be distinguished by \textit{what} they constrain: only the number of edges, the degrees of the agent nodes, the degrees of the artifact nodes, or the degrees of both the agent and artifact nodes. Second, ensembles can be distinguished by \textit{how} they impose these constraints: the constraints can be satisfied exactly, or only on average. In statistical physics, ensembles that impose exact or `hard' constraints are known as microcanonical, while ensembles that satisfy constraints on average or impose `soft' constraints are known as canonical.\cite{saracco2017inferring} 

Prior work on these ensembles generally adopts either a theoretical focus on the ensembles themselves, or an applied focus on the consequences of ensemble choice. In the theoretical literature, some (primarily mathematicians) have aimed to characterize the properties of ensembles, such as estimating the cardinality of the ensemble of matrices with fixed rows and columns (below, we call this ensemble $\mathcal{B}^{\text{FDSM}}$).\cite{barvinok2010number} Others (primarily physicists) have aimed to identify conditions under which ensembles are equivalent or non-equivalent, typically interpreting ensembles as representing thermodynamic systems.\cite{barre2007ensemble,touchette2015equivalence,squartini2015breaking} In the applied literature, the focus is not on identifying fundamental properties of ensembles, but instead on understanding the implications of choosing a particular ensemble when detecting a particular pattern, such as nestedness\cite{bruno2020ambiguity} or community structure.\cite{cimini2021meta,domagalski2021backbone} The present work falls into this latter group: we are not directly concerned with identifying fundamental properties of ensembles, but instead on identifying the consequences of ensemble choice, with the ultimate goal of offering practical guidance to applied researchers wishing to extract the backbone of a bipartite projection.

In the remaining subsections below, we first describe the FDSM in terms of its ensemble. We then present four potential alternative backbone models whose ensembles differ only slightly from FDSM, in terms of \textit{either} what they constrain or how they impose constraints. We then turn to exploring the consequences of choosing one of these alternatives over FDSM when extracting a backbone.

\subsubsection*{Fixed Degree Sequence Model (FDSM)}
In the \textit{fixed degree sequence model} (FDSM), $\mathbf{B^*}~\in~\mathcal{B}^{\text{FDSM}}$ are constrained to have the same agent and artifact degree sequences as $\mathbf{B}$. That is, FDSM constrains the degrees of both the agent and artifact nodes, and requires that these constraints are satisfied exactly, making it a tightly-constrained microcanonical ensemble. Adopting the FDSM implies, for example, that in all possible worlds a given species is found on exactly the same number of islands, and a given island is home to exactly the same number of species. The distribution of $P^*_{ij}$ arising from $\mathcal{B}^{\text{FDSM}}$ is unknown, but can be approximated by uniformly sampling $\mathbf{B^*}$ from $\mathcal{B}^\text{FDSM}$, constructing $\mathbf{P^*}$, and saving the values $P^*_{ij}$. In the studies below, we use 1000 samples of $\mathbf{B^*}$ generated using the `curveball' algorithm, which is among the fastest methods to sample $\mathcal{B}^\text{FDSM}$ uniformly at random.\cite{strona2014fast,Carstens_2015} The FDSM has been used to extract the backbone of bipartite projections of, for example, movies co-liked by viewers\cite{zweig2011systematic} and conference panel co-participation by scholars.\cite{stegbauer2012international,derudder2016international}

The FDSM offers an intuitively appealing approach to extracting the backbone of bipartite projections because it fully controls for both bipartite degree sequences, which are known to be responsible for many of the projection's structural characteristics.\cite{filho2020,guillaume2004bipartite} However, because the distribution of $P^*_{ij}$ must be computed via Monte Carlo sampling, it is computationally costly, making it impractical for all but relatively small bipartite projections. There are at least three distinct computational challenges. First, although the curveball algorithm is the fastest among existing methods for randomly sampling a bipartite graph with fixed degree sequences (i.e. for sampling $\mathbf{B^*}$ from $\mathcal{B}^\text{FDSM}$), it still can require several seconds per sample for large graphs. Second, once a $\mathbf{B^*}$ has been sampled, constructing each $\mathbf{P^*}$ requires matrix multiplication, which must be performed repeatedly and has complexity of at least $\mathcal{O}(n^{2.37})$.\cite{coppersmith1990} Finally, computing an edge's $p$-value (i.e. $\Pr(P^*_{ij} \geq P_{ij})$) with sufficient precision to achieve a specified familywise error rate that controls for Type-I error inflation due to multiple testing\cite{Tumminello} can require these sampling and multiplication steps to be performed a very large number of times (see \textit{Supplementary Text S2}).

These computational challenges have led researchers to develop other backbone models.\cite{neal2014backbone,saracco2015randomizing,saracco2017inferring} Many such models exist, however here we are focused on identifying methods that yield backbones similar to what would be obtained using FDSM, and thus which may serve as computationally-feasible alternatives to FDSM. Therefore, we consider only those models whose ensembles involve at least one of the two types of constraints imposed by FDSM. That is, we consider models that either (1) impose exact constraints, or (2) impose constraints on both the agent and artifact degrees.

\subsubsection*{Fixed Fill Model (FFM)}
In the \textit{fixed fill model} (FFM), $\mathbf{B^*}~\in~\mathcal{B}^{\text{FFM}}$ are simply constrained to contain the same number of $1$s as $\mathbf{B}$. That is, the FFM constrains only the number of edges, but requires that this constraint is satisfied exactly. Adopting the FFM implies, for example, that in all possible worlds only the total number of species-island pairs is fixed, but any given species may be found on a different number of islands and any given island may be home to a different number of species. The distribution of $P^*_{ij}$ arising from $\mathcal{B}^{\text{FFM}}$ has not been described before, but is derived in \textit{Supplementary Text S1.1}. We call it a \textit{Jacobi distribution} because it is related to Jacobi polynomials.

\subsubsection*{Fixed Row Model (FRM)}
In the \textit{fixed row model} (FRM), $\mathbf{B^*}~\in~\mathcal{B}^{\text{FRM}}$ are constrained to have the same agent degree sequence as $\mathbf{B}$, but have unconstrained artifact degree sequences. That is, the FRM constrains the degrees of the agent nodes, and requires that this constraint is satisfied exactly. A canonical variant of the FRM, the BiPCM$_r$, also constrains the degrees of the agent nodes, but only requires this constraint to be satisfied on average; we do not consider it here because it involves neither of FDSM's constraints.\cite{saracco2017inferring} Adopting the FRM for backbone extraction implies, for example, that in all possible worlds a given species is found on the same number of islands, but a given island may be home to a different number of species. The distribution of $P^*_{ij}$ arising from $\mathcal{B}^{\text{FRM}}$ is hypergeometric (see \textit{Supplementary Text S1.2}), and for this reason it is sometimes referred to as the hypergeometric model.\cite{Tumminello,neal2013identifying,cimini2021meta} The FRM has been used to extract the backbone of bipartite projections of, for example, movies co-starring actors,\cite{Tumminello} papers co-written by authors,\cite{Tumminello} parties co-attended by women,\cite{neal2013identifying} majority opinions joined by Supreme Court justices,\cite{neal2013identifying}, and microRNAs co-associated with diseases.\cite{chen2018bnpmda}

\subsubsection*{Fixed Column Model (FCM)}
In the \textit{fixed column model} (FCM), $\mathbf{B^*}~\in~\mathcal{B}^{\text{FCM}}$ are constrained to have the same artifact degree sequence as $\mathbf{B}$, but have unconstrained agent degree sequences. That is, the FCM constrains the degrees of the artifact nodes, and requires that this constraint is satisfied exactly. A canonical variant of the FCM, the BiPCM$_c$, also constrains the degrees of the artifact nodes, but only requires this constraint to be satisfied on average; we do not consider it here because it involves neither of FDSM's constraints.\cite{saracco2017inferring} Adopting the FCM for backbone extraction implies, for example, that in all possible worlds a given species may be found on a different number of islands, but a given island is home to the same number of species. The distribution of $P^*_{ij}$ arising from $\mathcal{B}^{\text{FCM}}$ has not been described before, but is derived in \textit{Supplementary Text S1.3}, where we show it is Poisson-binomial.

\subsubsection*{Stochastic Degree Sequence Model (SDSM)}
Finally, the \textit{stochastic degree sequence model} (SDSM) takes $\mathcal{B}^{\text{SDSM}}$ to be all binary $m \times n$ matrices, but also gives a process for generating these matrices with different probabilities. Each $\mathbf{B^*}$ is generated by filling the cells $B^*_{ik}$ with a $0$ or $1$ depending on the outcome of an independent Bernoulli trial with probability $p^*_{ik}$. The distribution of the random variable $P^*_{ij}$ arising from $\mathcal{B}^{\text{SDSM}}$ is Poisson-binomial with parameters which can be computed using the $p^*_{ik}$ (see \textit{Supplementary Text S1.4}).\cite{domagalski2021backbone,liebig2016} There are many ways to choose $p^*_{ik}$, but in the studies below we choose $p^*_{ik}$ so that it approximates $\Pr(B^*_{ik} = 1)$ for $\mathbf{B^*}~\in~\mathcal{B}^{\text{FDSM}}$. This choice of $p^*_{ik}$ ensures that the SDSM constrains the degrees of both the agent and artifact nodes, but only requires these constraints to be satisfied on average. Adopting such a version of SDSM implies, for example, that in each possible world a given species may be found on many or few islands and a given island may be home to many or few species, but the \textit{average} number of islands on which a given species lives in all possible worlds and the \textit{average} number of species that live on an given island in all possible worlds matches these values the observed world. The SDSM has been used to extract the backbone of bipartite projections of, for example, legislators co-sponsoring bills, \cite{neal2020,neal2014backbone,schoch2020legislators,aref2020detecting,aref2021detecting} zebrafish (\textit{Danio rerio}) sharing operational taxonomic units,\cite{buerger2020gastrointestinal} countries sharing exports,\cite{saracco2015randomizing} and genes expressed in genesets.\cite{marini2021genetonic}

\section*{Study 1: Choosing cell-filling probabilities for the SDSM}
The SDSM requires choosing $p^*_{ik}$, which we want to approximate $\Pr(B^*_{ik} = 1)$ for $\mathbf{B^*}~\in~\mathcal{B}^{\text{FDSM}}$. There are three types of methods that might be used for doing so: arithmetic, general linear models, and entropy maximization. First, we can choose $p^*_{ik} = (r_i~\times~c_k)/f$, where $r_i$ is the sum of entries in row $i$ of $\mathbf{B}$, $c_k$ is the sum of entries in column $k$ of $\mathbf{B}$, and $f$ is the sum of all entries in $\mathbf{B}$. When $p^*_{ik}$ falls outside the $[0,1]$ range, it is simply truncated toward $0$ or $1$, respectively. This method has a long history in ecology;\cite{gotelli2000null} we call it RCF because the value is chosen based on a row sum, a column sum, and the number of entries of $\mathbf{B}$ that are filled with a one, but elsewhere it has been called the `Chung-Lu method'.\cite{becatti2019entropy,chung2002connected} Second, an estimate can be obtained by fitting a general linear model of the form:
\begin{align*}
B_{ik} &= \beta_0 + \beta_1r_i + \beta_2c_k + \epsilon \text{, or} \\
B_{ik} &= \beta_0 + \beta_1r_i + \beta_2c_k + \beta_3r_ic_k + \epsilon,
\end{align*}
where the $\beta$'s are estimated coefficients and $\epsilon$ is an error term. If the model is treated as a linear regression and the coefficients are estimated using ordinary least squares, then the predicted value of $B_{ik}$ is chosen for $p^*_{ik}$, either truncating values outside the required $[0,1]$ range (linear probability model; LPM) or transforming them into the required range using a linear discriminant model (LDM).\cite{allison2020better} If the model is treated as a logistic regression and the coefficients are estimated using maximum likelihood, then the predicted probability that $B_{ik} = 1$ is chosen for $p^*_{ik}$. In prior work, the logistic regression approach has used a scobit or logit link function, with or without an interaction term ($\beta_3$).\cite{neal2014backbone,schoch2020legislators,neal2020} Finally, an estimate can be obtained by entropy maximization methods, including the polytope method (Poly)\cite{domagalski2021backbone,neal_domagalski_yan} or bipartite configuration model (BiCM).\cite{saracco2017inferring,saracco2015randomizing,pythonbicm} In this study, we evaluate the accuracy and speed of these methods for choosing $p^*_{ik}$ that approximate $\Pr(B^*_{ik} = 1)$ for $\mathbf{B^*}~\in~\mathcal{B}^{\text{FDSM}}$.

\subsection*{Methods}
To evaluate accuracy, we begin by enumerating all the members of a small $\mathcal{B}^\text{FDSM}$. For example, given an agent degree sequence of $[1,1,2]$ and an artifact degree sequence of $[1,1,2]$, $\mathcal{B}^\text{FDSM}$ contains 5 members (see Table~\ref{tab:probabilities}A). Second, from this complete enumeration, we compute the probabilities we wish $p^*_{ik}$ to approximate (i.e., $\Pr(B^*_{ik} = 1)$ for $\mathbf{B^*}~\in~\mathcal{B}^{\text{FDSM}}$, see Table~\ref{tab:probabilities}B). Third, we compute $p^*_{ik}$ using each of nine methods (see Table~\ref{tab:probabilities}C for values obtained using the BiCM method). Finally, we quantify the accuracy with which $p^*_{ik}$ approximates the desired probabilities using the absolute mean difference for all $i,k$. In the example shown in Table~\ref{tab:probabilities}, BiCM's accuracy for these degree sequences is 0.028. That is, on average $p^*_{ik}$ chosen using BiCM deviates from the desired probabilities by $\pm$ 0.028 on average. Because evaluating accuracy in this way requires enumerating all members of $\mathcal{B}^\text{FDSM}$, it is possible only for short degree sequences that define $\mathcal{B}^\text{FDSM}$ with small cardinality. We focus on degree sequences ranging in length from $2$ to $5$, which define 384 unique $\mathcal{B}^\text{FDSM}$ ranging in cardinality from 4 to 2040.

\begin{table}[h]
\centering
\begin{tabular}{lcr}
\multicolumn{3}{c}{(\textbf{A}) Members of $\mathcal{B}^\text{FDSM}$} \\
\multicolumn{3}{c}{
\begin{tabular}{ccccccccc}
\begin{tabular}{|c|c|c|}
\hline
1 & 0 & 0 \\
\hline
0 & 0 & 1 \\
\hline
0 & 1 & 1 \\
\hline
\end{tabular}
&
\begin{tabular}{|c|c|c|}
\hline
0 & 0 & 1 \\
\hline
1 & 0 & 0 \\
\hline
0 & 1 & 1 \\
\hline
\end{tabular}
&
\begin{tabular}{|c|c|c|}
\hline
0 & 0 & 1 \\
\hline
0 & 0 & 1 \\
\hline
1 & 1 & 0 \\
\hline
\end{tabular}
&
\begin{tabular}{|c|c|c|}
\hline
0 & 0 & 1 \\
\hline
0 & 1 & 0 \\
\hline
1 & 0 & 1 \\
\hline
\end{tabular}
&
\begin{tabular}{|c|c|c|}
\hline
0 & 1 & 0 \\
\hline
0 & 0 & 1 \\
\hline
1 & 0 & 1 \\
\hline
\end{tabular}
\end{tabular}
}\\
& & \\
(\textbf{B}) Desired probabilities & & (\textbf{C}) $p^*_{ik}$ computed using BiCM\\
\begin{tabular}{|c|c|c|}
\hline
0.2 & 0.2 & 0.6 \\
\hline
0.2 & 0.2 & 0.6 \\
\hline
0.6 & 0.6 & 0.8 \\
\hline
\end{tabular}
&
&
\begin{tabular}{|c|c|c|}
\hline
0.216 & 0.216 & 0.568 \\
\hline
0.216 & 0.216 & 0.568 \\
\hline
0.568 & 0.568 & 0.863 \\
\hline
\end{tabular}\\
 & & \\
\end{tabular}
\caption{SDSM probabilities given agent and artifact degree sequences [1,1,2]}
\label{tab:probabilities}
\end{table}

After identifying each method's accuracy, we evaluate the computational running time of the four most accurate methods by using them to choose $p^*_{ik}$ for bipartite graphs defined by up to 1000 agents and up to 1000 artifacts, and thus requiring choosing up to 1,000,000 probabilities.

\subsection*{Results}
\begin{figure}
    \centering
    \includegraphics[width=.8\textwidth]{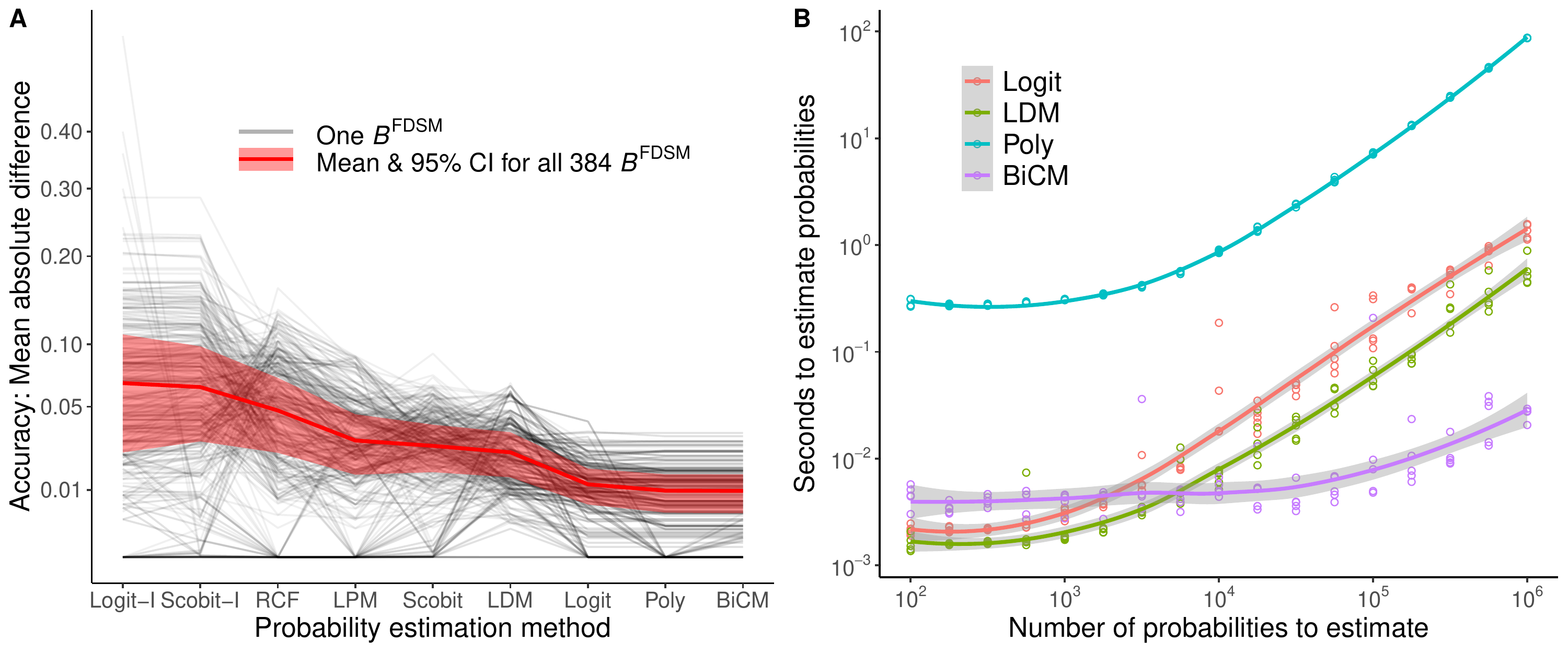}
    \caption{(A) Accuracy and (B) speed computing $p^*_{ik}$ using different methods. Lines show means, while shaded regions show 95\% confidence intervals.}
    \label{fig:study1}
\end{figure}

Figure~\ref{fig:study1}A shows the accuracy of each method's computation of $p^*_{ik}$. Each gray line plots the accuracy of each method for a single $\mathcal{B}^\text{FDSM}$, while the red line and shaded region plots the mean and 95\% confidence interval of the accuracy of each method over all 384 $\mathcal{B}^\text{FDSM}$. We find that choosing $p^*_{ik}$ using a logistic regression with an interaction term (i.e., Scobit-I and Logit-I) is on average least accurate,\cite{neal2014backbone,neal2020} while choosing $p^*_{ik}$ using the two entropy maximization method (BiCM and Poly) yield numerically equivalent results, which were on average most accurate.\cite{domagalski2021backbone,saracco2015randomizing}

Figure~\ref{fig:study1}B shows the number of seconds required to compute $p^*_{ik}$ using a 2.3 GhZ Intel i7 processor; lines illustrate the mean running time, while the shaded regions show the 95\% confidence interval. Among the two most accurate methods, BiCM is several orders of magnitude faster than Polytope. When computing more than $10^4$ probabilities, BiCM is also faster than the two slightly less accurate Logit and LDM methods. In the largest case we evaluated, computing $10^6$ probabilities, BiCM took only about 0.026 seconds. Therefore, we use BiCM for choosing $p^*_{ik}$ when extracting SDSM backbones in the remaining studies because it is both the most accurate and fastest.

\section*{Study 2: Statistical power of SDSM}
Ensemble backbone models require the specification of a statistical significance level $\alpha$, which determines how uncommonly large an observed edge weight $P_{ij}$ must be when compared to edge weights $P^*_{ij}$ arising from an ensemble in order for a corresponding edge to be included in the backbone. For a given model, smaller values of $\alpha$ represent more stringent criteria for retaining edges, and therefore yield sparser backbones. Although FDSM and SDSM define their respective ensembles by constraining both agent and artifact degree sequences, and thus aim to yield similar backbones, a given $\alpha$ does not necessarily represent the same level of stringency in these two models. Because the SDSM allows variation in the degree sequences of $\mathbf{B^*}~\in~\mathcal{B}^\text{SDSM}$, the distribution of $P^*_{ij}$ is wider.\cite{cimini2021meta,neal2021} These wider distributions mean that the SDSM provides a more conservative test of edge weight significance than FDSM, or alternatively the SDSM has less statistical power to detect significant edges than FDSM. 

A concrete example serves to illustrate this difference. In economic geography, it is common to study the world city network using a bipartite projection where two cities are linked to the extent that firms maintain locations in both cities. The Globalization and World Cities (GaWC) dataset has been widely-used in this context, and takes the form of a bipartite network recording the presence or absence of 100 firms (artifacts) in 196 cities (agents) in the year 2000.\cite{taylor2002measurement,neal2021} In this bipartite network, the agent degrees are right-tailed because most cities contain only a few firms, while a few cities such as New York contain many. Likewise, the artifact degrees are also right tailed because most firms maintain locations in only a few cities, while a few firms such as the accounting firm KPMG maintain locations in many.

Figure \ref{fig:study2}A illustrates the distribution of the Milan-Paris edge weight in projections arising from $\mathcal{B}^\text{FDSM}$ and $\mathcal{B}^\text{SDSM}$ of which the observed bipartite network is a member (i.e., the random variable $P^*_{ij})$. These distributions allow a researcher to decide whether Milan and Paris's observed number of co-located firms is significantly large, and therefore whether Milan and Paris should be connected in a world city network backbone. The SDSM distribution is wider than the FDSM distribution,\cite{cimini2021meta,neal2021} which has implications for whether the Milan-Paris edge will be included in a backbone extracted at a given significance level using each model. In the observed data, there are 26 firms co-located in Milan and Paris (i.e., $P_{ij} = 26$). The probability of observing the same or larger edge weight in projections from the FDSM ensemble is 0.0033, which is less than $\frac{0.05}{2}$, and therefore a Milan-Paris edge is deemed significant by the FDSM and is included in the FDSM backbone extracted at $\alpha = 0.05$. In contrast, the probability of observing the same or larger edge weight in projections from the SDSM ensemble is 0.0275, which is \textit{not} less than $\frac{0.05}{2}$, and therefore a Milan-Paris edge is \textit{not} deemed significant by the SDSM and is \textit{not} included in the SDSM backbone extracted at $\alpha = 0.05$. For a given level of significance $\alpha$, this difference in statistical power leads the SDSM backbone to be sparser than the FDSM backbone (density = 0.004 vs. 0.012), and means that these two backbones are dissimilar (Jaccard = 0.36).

In this study, we investigate SDSM's statistical power relative to FDSM, and specifically whether extracting an SDSM backbone using a more liberal (i.e., larger) $\alpha$ makes it more similar to an FDSM backbone extracted at $\alpha = 0.05$.

\subsection*{Methods}
To evaluate SDSM's statistical power and the effect of significance levels on the similarity of SDSM and FDSM backbones, we first extracted the FDSM backbone from the GaWC bipartite network at $\alpha = 0.05$. We then extracted SDSM backbones from the GaWC bipartite network at $0.01 \leq \alpha \leq 0.3$ in $0.001$ increments, each time computing the Jaccard index ($J$) to measure the similarity between the SDSM and FDSM backbones. After comparing SDSM and FDSM backbones extracted from the empirical GaWC bipartite network, we repeat this process using 100 synthetic bipartite networks with the same dimensions ($196 \times 100$), density ($0.08$) and right-tailed agent and artifact degree distributions.

\subsection*{Results}
\begin{figure}
    \centering
    \includegraphics[width=.8\textwidth]{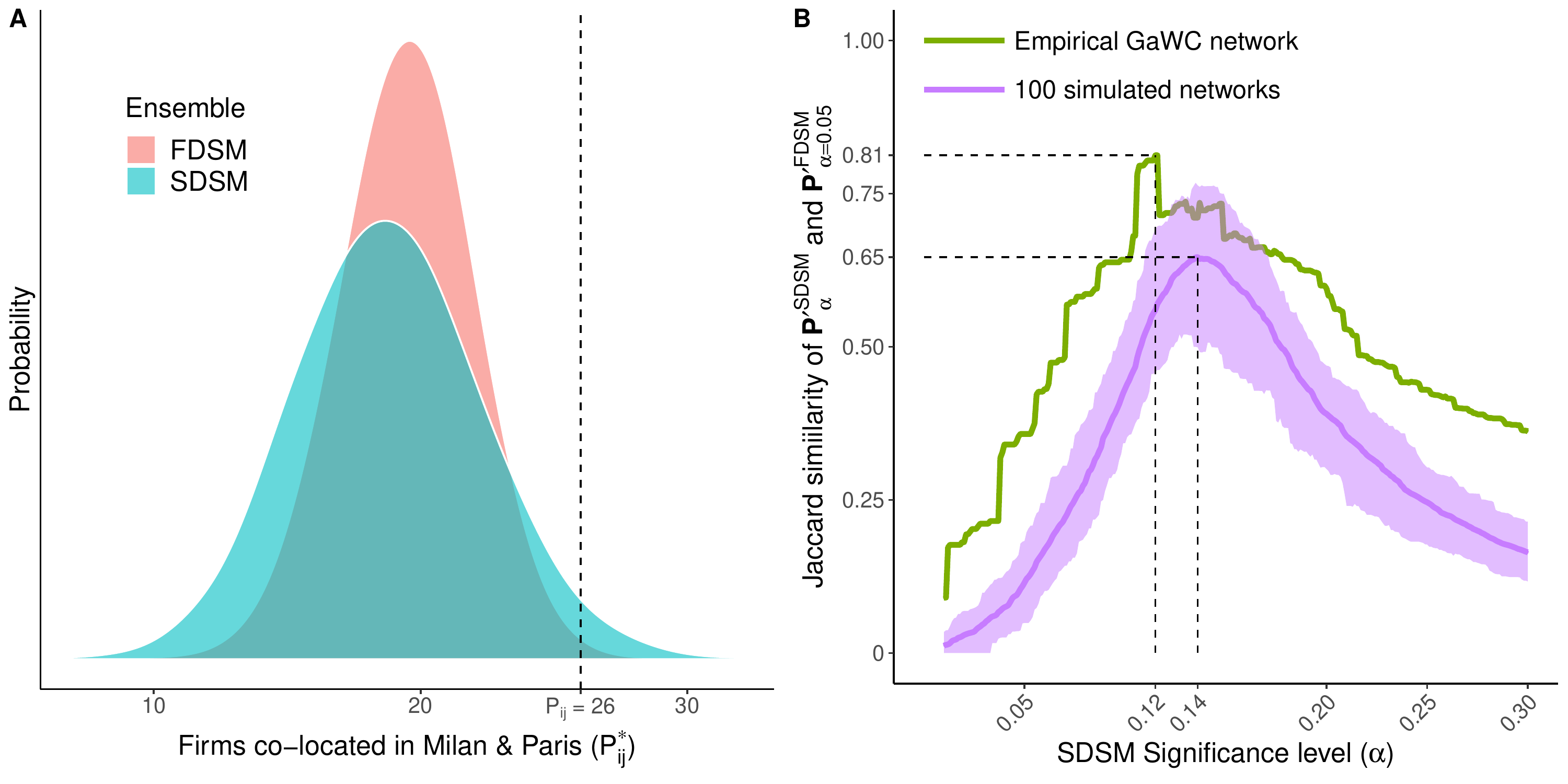}
    \caption{Statistical power of SDSM. (A) Distribution of weights for the Paris-Milan edge in projections derived from FDSM and SDSM ensembles. (B) Similarity of an FDSM backbone extracted at $\alpha = 0.05$ to SDSM backbones extracted at various $\alpha$ from an empirical bipartite network (green line) and from 100 synthetic bipartite networks (purple line = mean, purple region = 10\textsuperscript{th}--90\textsuperscript{th} percentile).}
    \label{fig:study2}
\end{figure}

The green line in Figure \ref{fig:study2}B shows the Jaccard similarity between an FDSM backbone extracted from the empirical GaWC network at $\alpha = 0.05$ and SDSM backbones extracted at the significance levels shown on the x-axis. We find that an SDSM backbone achieves its maximum similarity to the FDSM backbone ($J = 0.81$) when it is extracted using the more liberal significance level of $\alpha = 0.12$. Returning to the example in Figure \ref{fig:study2}A, using this more liberal significance level would result in the Milan-Paris edge being deemed significant and included in the SDSM backbone because its SDSM $p$-value $0.0275 < \frac{0.12}{2}$. Because this more liberal significance level results in the inclusion of additional edges, the new SDSM backbone extracted at $\alpha = 0.12$ has a density ($0.01$), which is closer to that of the FDSM backbone extracted at $\alpha = 0.05$ ($0.012$).

The purple line in Figure \ref{fig:study2}B shows the mean Jaccard similarity between an FDSM backbone extracted using $\alpha = 0.05$ and SDSM backbones extracted using $0.01 \leq \alpha \leq 0.3$ from 100 bipartite networks generated to resemble the empirical GaWC network. The shaded purple region shows the 10\textsuperscript{th} and 90\textsuperscript{th} percentile of Jaccard similarities of these backbones. We find that these synthetic networks behave similarly to the empirical network. Specifically, SDSM and FDSM backbones extracted from a low-density $196 \times 100$ bipartite network with right-tailed degree distributions achieve a maximum similarity of $0.49 < J < 0.76$ when the FDSM backbone is extracted using $\alpha = 0.05$ and the SDSM backbone is extracted using $\alpha = 0.14$. This is promising because it suggests that, given the characteristics of an empirical bipartite network, it may be possible to select a significance level for extracting a computationally-efficient SDSM backbone that closely resembles a computationally-infeasible FDSM backbone.

\section*{Study 3: Backbone similarity under varying degree distributions}
Agent and artifact degree distributions are a key feature of a bipartite network, and are known to have implications for bipartite projections. \cite{filho2020,domagalski2021backbone,neal2021} The FDSM is particularly appealing because it allows decisions about the significance of edges in a projection to be conditioned on both bipartite degree sequences, thereby taking into account these important features. However, because the computational requirements of the FDSM make it impractical for extracting the backbone from most bipartite projections, it is often necessary to use a different backbone model. In this study, we evaluate the similarity of an FDSM backbone and backbones extracted using more computationally efficient models. We perform this comparison for backbones extracted from bipartite networks characterized by five types of degree distributions: right-tailed, left-tailed, normal, constant, and uniform. 

For the sake of concreteness, in this section we use the example of a bipartite network in which authors (agents) are linked to the papers they have written (artifacts). The projection of such a network yields a co-authorship network in which the edge weight between a pair of authors indicates their number of co-authored papers.\cite{newman2001} These edge weight values will depend heavily on the distribution of papers written by authors (i.e., the agent degree sequence), and on the distribution of authors on each paper (i.e., the artifact degree sequence). Different degree distributions describe different kinds of scholarly environments as shown in Table \ref{tab:distributions}.  The choice of a backbone model affects whether these distributions are considered, and in this example affects whether decisions about the significance of two authors' number of co-authored papers consider the scholarly environment. The FDSM compares their observed number of co-authored papers to the number that might be observed in alternative realizations \textit{of the same environment}, while other backbone models relax the extent to which the environment is held constant.

\begin{table}[]
\begin{tabular}{p{.23\textwidth}p{.35\textwidth}p{.35\textwidth}}
\hline
\textbf{Degree Distribution} & \textbf{Authors (agents)} & \textbf{Papers (artifacts)} \\
\hline
Right-tailed $\sim\beta(1,10$) & Most write some papers, but a few are prolific (most departments). & Most papers are sole-authored, but some are written by large teams (e.g., sociology). \\
Left-tailed $\sim\beta(10,1$)& Most are prolific, but some are inactive (elite departments). & Most papers are written by large teams, but some are sole-authored (e.g., physics). \\
Uniform $\sim\beta(1,1$) & There is substantial diversity in scholarly output (e.g., interdisciplinary departments). & There is substantial diversity in the size of authorship teams (e.g., an entire university). \\
Constant $\sim\beta(10000,10000$) & There are strong norms about how many papers an author should have (e.g., for performance evaluations). & There are strong norms about how many authors a paper should have (e.g., two: a senior author \& a junior author) \\
Normal $\sim\beta(10,10$) & Scholarly output varies around some typical level. & Authorship teams vary around some typical size. \\
\hline
\end{tabular}
\caption{Bipartite degree distributions, with examples in the context of a scholarly authorship bipartite network}
\label{tab:distributions}
\end{table}

\subsection*{Methods}
We evaluate similarities among the backbones extracted using different models by comparing backbones extracted from synthetic $100 \times 100$ bipartite networks with a density of 0.1, and with a combination of agent and artifact degree distributions shown in Table \ref{tab:distributions}. Following our example, these synthetic bipartite networks might represent a college of 100 faculty who collectively wrote 100 papers, in a particular type of scholarly environment where each individual had a 10\% chance of being an author on each paper. After generating a bipartite network with a given size, density, and degree distributions, we extract five different backbones from the generated bipartite network, using the fixed fill model, fixed row model, fixed column model, stochastic degree sequence model, and fixed degree sequence model; in all cases we use $\alpha = 0.05$. We compute the similarity of the first four backbones to the FDSM backbone using a Jaccard index, repeating this process 100 times for each of the 25 possible combinations of agent and artifact degree distributions.

\subsection*{Results}
\begin{figure}
    \centering
    \includegraphics[width=.8\textwidth]{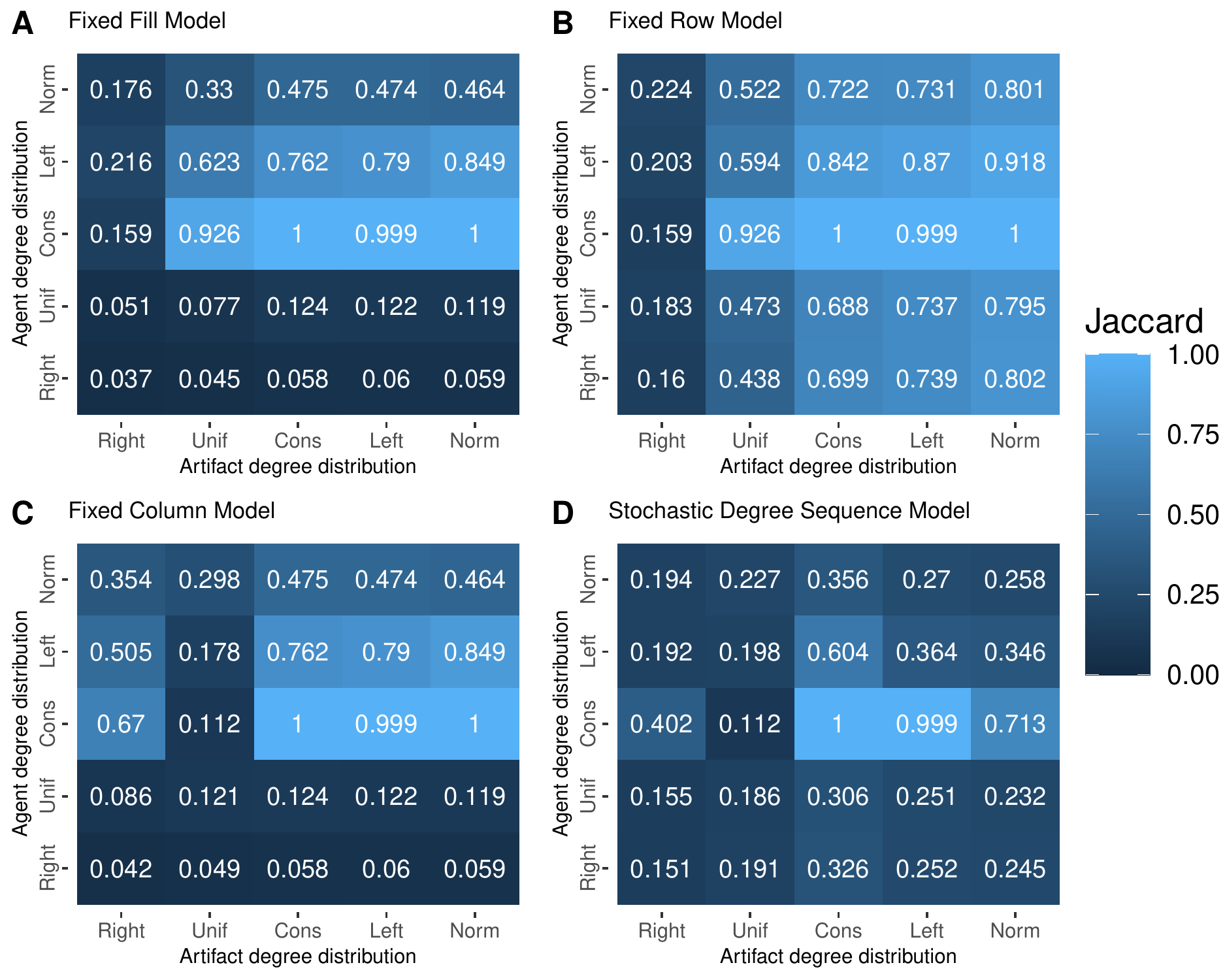}
    \caption{Jaccard similarity of a backbone extracted at $\alpha = 0.05$ using the Fixed Degree Sequence Model and a backbone extracted using (A) the Fixed Fill Model, (B) Fixed Row Model, (C) Fixed Column Model, (D) Stochastic Degree Sequence Model. Each cell represents the mean over 100 instances of a $100 \times 100$ bipartite network with given agent and artifact degree distributions.}
    \label{fig:study3}
\end{figure}

The heatmaps in Figure \ref{fig:study3} illustrate the similarity between an FDSM backbone and a backbone extracted using an alternative model. The rows of each heat map correspond to different agent degree distributions, and the columns correspond to different artifact degree distributions, in the synthetic bipartite networks from which the backbones were extracted. The lightest patches identify conditions under which a given backbone model yields a backbone that is similar to what would be obtained using the computationally costly FDSM, while darker patches identify conditions under which these two backbones differ. We find that when agent degrees are constant (i.e., every agent has the same degree) and artifact degrees are constant or left-tailed, all backbone models yield the same backbone as FDSM (Mean $J = 1$). However, beyond this special case, which is likely to be rare in empirical data, similarity to FDSM-extracted backbones varies.

As expected, the similarity of backbones extracted using FRM and FDSM depends primarily on the distribution of artifact degrees, not agent degrees (see Figure \ref{fig:study3}B). For example, for any agent degree distribution, these two models yield very different backbones when artifact degrees follow a right-tailed distribution (Mean $J = 0.186$), but very similar backbones when artifact degrees follow a normal distribution (Mean $J = 0.863$). This occurs because both models exactly control for agent degrees, however FDSM also controls for artifact degrees, while FRM does not. 

A similar but rotated pattern emerges when considering the FCM: the similarity of backbones extracted using FCM and FDSM depends primarily on the distribution of agent degrees, not artifact degrees (see Figure \ref{fig:study3}C). For any artifact degree distribution, these two models yield very different backbones when agent degrees follow a right-tailed or uniform (Mean $J = 0.084$) distribution , but more similar backbones when agent degrees follow a left-tailed distribution or are constant (Mean $J = 0.617$). This occurs because both models exactly control for artifact degrees, however FDSM also controls for agent degrees, while FRM does not. However, there is a notable exception to this general pattern: when artifact degrees follow a uniform distribution, FCM and FDSM always yield different backbones (Mean $J = 0.151$).

The conditions under which the FFM yields FDSM-similar backbones occur at the intersection of the conditions under which the FRM and FCM both yield FDSM-like backbones (see Figure \ref{fig:study3}A). When artifact degrees follow a right-tailed distribution or the agent degrees follow a right-tailed or uniform distribution, then FFM and FDSM backbones differ (Mean $J = 0.1$). In contrast, for other combinations of degree distributions, FFM and FDSM backbones are more similar (Mean $J = 0.724$).

Finally, as expected based on the findings from study 2, we observe that the SDSM generally yields different backbones than FDSM when both are extracted at $\alpha = 0.05$ (see Figure \ref{fig:study3}D). Specifically, except in the narrow case where agent degrees are constant and artifact degrees are constant or left-tailed (Mean $J = 1$), SDSM and FDSM backbones exhibit only modest similarity (Mean $J = 0.314$). This lack of similarity occurs because SDSM offers a less statistically powerful (or more conservative) test of edges statistical significance than FDSM, and therefore retains fewer edges in the backbone. However, findings from study 2 also suggested that careful selection of the significance level used for extracting an SDSM backbone can yield results more similar to FDSM.

\begin{figure}
    \centering
    \includegraphics[width=.8\textwidth]{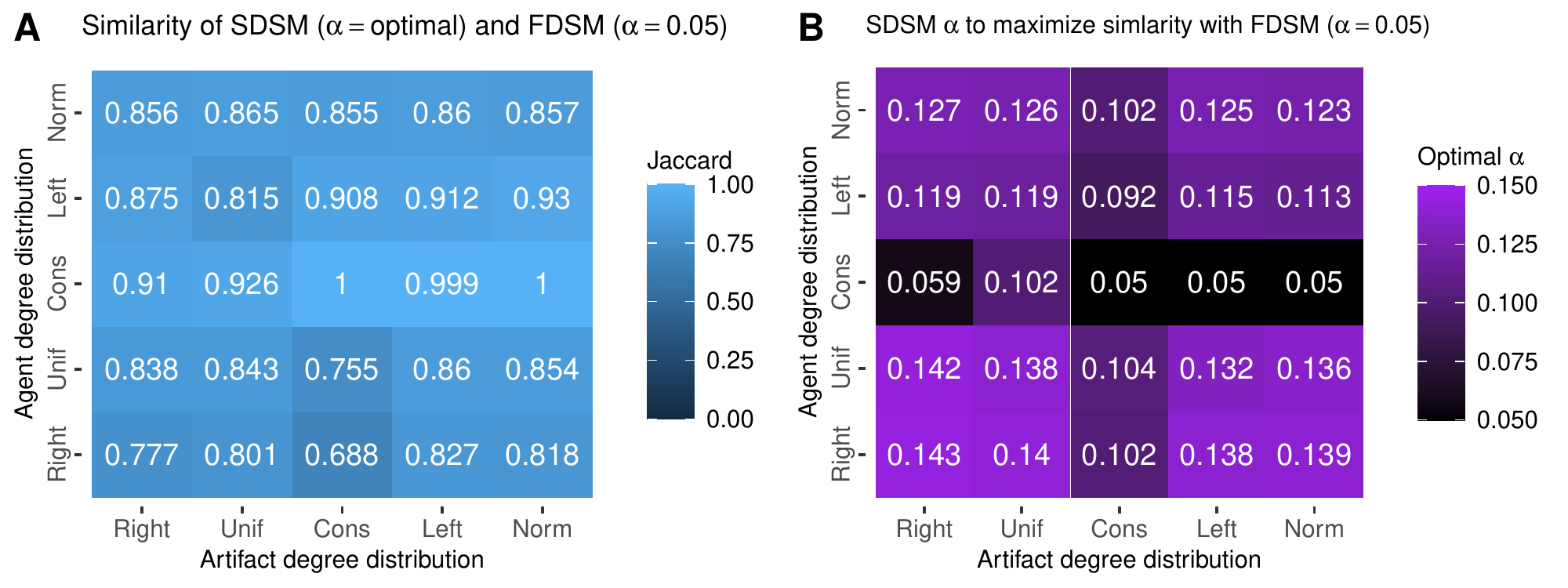}
    \caption{(A) Given agent and artifact degree distributions, there exists a statistical significance level $\alpha$ that maximizes the similarity between an SDSM backbone extracted at this level and an FDSM backbone extracted at $\alpha = 0.05$, and (B) when used yields an SDSM backbone that is very similar to the corresponding FDSM backbone.}
    \label{fig:study3_optim}
\end{figure}

To explore this possibility, we expanded the analysis reported in figure \ref{fig:study3}D by extracting SDSM backbones at different significance levels $\alpha$. We find that when a suitably more liberal (i.e., larger) significance level $\alpha$ is used to extract an SDSM backbone, the resulting SDSM backbone is very similar to an FDSM backbone extracted at $\alpha = 0.05$ (see Figure \ref{fig:study3_optim}A). Specifically, for backbones extracted from bipartite networks with \textit{any} agent or artifact degree distributions, these two backbones tend to be very similar (Mean $J = 0.865$). This suggests that in principle the fast SDSM can be used to obtain a close approximation of a computationally-infeasible FDSM backbone from any bipartite network.

In practice, using SDSM to obtain an FDSM-like backbone requires selecting an $\alpha$ value for the SDSM that corresponds to $\alpha = 0.05$ in the FDSM. We observe that there are three distinct values of such an `optimal' $\alpha$ that depend on agent and artifact degree distributions (see Figure \ref{fig:study3_optim}B). First, when agent degrees are constant, a value only slightly higher than $0.05$ (Mean $= 0.062$, SD $= 0.021$) achieves the best approximation of an FDSM backbone. Second, when artifact degrees are constant, a value roughly double (Mean $= 0.09$, SD $= 0.022$) achieves the best approximation of an FDSM backbone. Finally, when neither agent nor artifact degrees are constant, which is likely in most empirical bipartite networks, a value roughly 2.5 times larger (Mean $= 0.13$, SD $= 0.014$) achieves the best approximation of an FDSM backbone. Although further work is needed to facilitate the \textit{a priori} selection of an $\alpha$ that allows an SDSM backbone to closely approximate an FDSM$_{\alpha = 0.05}$ backbone, these results suggest that under the most common circumstances (i.e., when there is variation in degrees) $\alpha \approx 0.13$ may be appropriate.

\section*{Study 4: Recovery of community structure}
Studies 1-3 examine the backbones extracted from random bipartite networks; however, empirical bipartite networks are not random. Frequently they contain a block structure that implies a particular community structure in the bipartite projection. In this study, we evaluate the extent to which backbones extracted using different models reflect a known community structure that is encoded in the bipartite data from which they are extracted.\cite{cann2018correct} Recent work has shown that FDSM, FRM, SDSM, and BiPCM (a canonical variant of FRM) each yield backbones with \textit{similar} communities structures.\cite{cimini2021meta} Other work has shown that SDSM and FDSM backbones extracted from a bipartite network representing bill co-sponsorship in the 114\textsuperscript{th} session of the US Senate more clearly captured the \textit{hypothesized} partisan community structure than an FRM backbone.\cite{domagalski2021backbone} We build on this prior work using synthetic data that is constructed to contain a ground truth communities, which allows us to evaluate backbone models' ability to recover \textit{true} communities, and not simply similar or hypothesized ones.

\subsection*{Methods}
We investigate the ability for backbones to recover a known community structure in three steps. First, we simulate a $200 \times 1000$ bipartite network with a density of 0.1 and right-tailed agent and artifact degree distributions. We focus on a bipartite network with more artifacts than agents to ensure that these data contain sufficient information to encode potential community memberships. We focus on a bipartite network with right-tailed degree distributions because they are common in many empirical unipartite\cite{broido2019scale} and bipartite networks.\cite{neal2020,neal2021,ahn2011flavor} This synthetic bipartite network could represent a legislative body composed of 200 legislators casting votes on 1000 bills, where any given legislator had a 10\% chance of voting in favor of any given bill. The right-tailed degree distributions capture the fact that most legislators vote in favor of only a few bills, and that most bills receive the support of only a few legislators, which is typical of legislative bodies. The backbone of a projection of such a bipartite network would represent a network of collaboration or ideological alignment among legislators.\cite{neal2020}

\begin{figure}
    \centering
    \includegraphics[width=.8\textwidth]{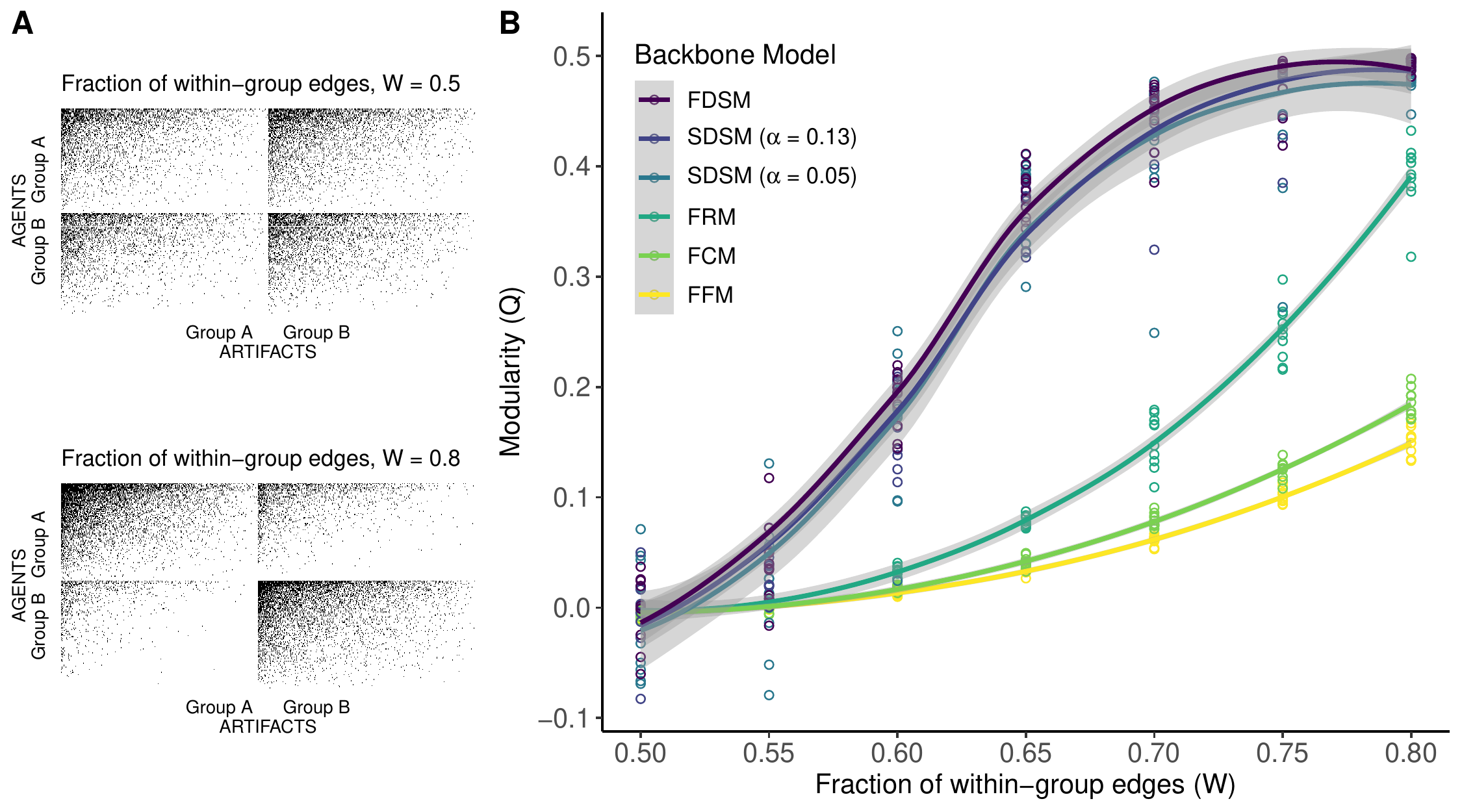}
    \caption{(A) Synthetic bipartite networks with varying levels of block structure, from which (B) backbones extracted using different models exhibit varying modularity.}
    \label{fig:study4}
\end{figure}

Second, we incorporate evidence of communities in this bipartite network by randomly assigning each agent and each artifact to one of two groups. We then perform checkerboard swaps, which preserve the degree distributions, until a given fraction of edges $W$ are within-group, connecting an agent and artifact from the same group.\cite{guimera2007module} Figure \ref{fig:study4}A provides graphical depictions of the matrices describing synthetic bipartite networks at two values of $W$. In each plot, the rows represent agents assigned to group A or B, the columns represent artifacts assigned to group A or B, and a cell is shaded black if the row agent is connected to the column artifact. When $W = 0.5$, agents in a given group are equally likely to associate with artifacts in either group, placing $\approx 0.5$ of the edges (i.e., shaded cells) in the diagonal blocks and $\approx 0.5$ of the edges in the off-diagonal blocks. In contrast, when $W = 0.8$, agents in a given group are much more likely to associate with artifacts from their own group than artifacts in the other group, placing $\approx 0.8$ of the edges in the diagonal blocks and $\approx 0.2$ of the edges in the off-diagonal blocks. Returning to our example, the groups could represent political parties: each legislator belongs to one of two parties (i.e., there are conservative and liberal legislators), and each bill advances the agenda of one of these parties (i.e., there are conservative and liberal bills). When $W = 0.5$, a conservative legislator is equally likely to vote for conservative and liberal bills, while when $W = 0.8$, a conservative legislator is four-times more likely to vote for a conservative bill than a liberal bill.

Finally, we extract a backbone from the bipartite network using a given model and compute the backbone's modularity $Q$ with respect to the agents' group assignments.\cite{newman2004finding} If a backbone model is able to recover the community structure from evidence in the bipartite network, then we expect a positive association between $W$ and $Q$. In the legislative example, if legislators are bipartisan in their voting patterns (i.e., $W = 0.5$), then legislators should not be clustered by party in the backbone (i.e., $Q \approx 0$). In contrast, if legislators are strongly partisan in their voting patterns (i.e., $W = 0.8$), then legislators should be clustered by party in the backbone (i.e., $Q \gg 0$).

We repeat these three steps $10$ times for $0.5 \leq W \leq 0.8$ in $0.05$ increments. When evaluating the SDSM backbone, we consider both a backbone extracted using the conventional significance level of $\alpha = 0.05$ and one extracted at the more liberal $\alpha = 0.13$, which study 3 suggests yields a backbone similar to FDSM.

\subsection*{Results}
Figure \ref{fig:study4}B shows the modularity (y-axis; with respect to known community memberships) of backbones extracted using different models from bipartite networks containing different fractions of within-community edges (x-axis). Solid lines illustrate the mean modularity across 10 replications, while the shaded regions illustrate 95\% confidence intervals. All six lines increase monotonically, confirming that all backbone models yield backbones that can recover a known community structure; however, there is notable variation among the models. As evidence of community structure grows stronger in the bipartite network, the modularity of backbones extracted using the FFM and FCM slowly increase, but even when the evidence of such a structure is quite strong (i.e., when $W = 0.8$) they only achieve average values of $Q = 0.15$ and $0.18$, respectively. Backbones extracted using the FRM display a similar pattern, but achieve a statistically significantly higher average modularity ($Q = 0.39$) value when $W$ is large. 

In contrast, backbones extracted using FDSM and SDSM yield modularity values that are statistically significantly larger than those obtained from FFM, FRM, or FCM backbones, but that are not statistically significantly different from each other. That is, these backbone models are indistinguishable in their ability to recover the known community structure, and do so very well. As evidence of a community structure grows stronger in the bipartite network, the modularity of backbones extracted using these models rapidly increases. When the evidence of community structure is strong (i.e., when $W = 0.8)$, these backbones have very high modularity (mean $Q = 0.49$). However, even when there is only modest evidence of community structure in the bipartite network (e.g., when $W = 0.65$), these backbones are still able to identify the community structure and have a distinctively high modularity (mean $Q = 0.37$).

These findings suggest that although all backbone models can yield backbones that recover a known community structure, SDSM and FDSM backbones are able to detect this structure more clearly and from a weaker signal.

\section*{Discussion} 
Bipartite networks can be used to represent a wide range of phenomena in the social and natural worlds including interspecies competition, global trade, scientific advances, and legislative deliberation. Likewise, projections of bipartite networks, which take the form of co-occurrence networks, can be useful for inferring unipartite networks whose edges would otherwise be difficult to measure directly. The fixed degree sequence model (FDSM) offers an appealing null model for making such inferences, but its computational complexity often makes it impractical. Several computationally simpler alternatives to FDSM have been, including the fixed fill model (FFM) fixed row model (FRM), fixed column model (FCM), and stochastic degree sequence model (SDSM). In this paper we have systematically compared FDSM to each of these alternatives to evaluate their aspects of their accuracy, speed, statistical power, backbone similarity, and ability to recover a known community structure.

In study 1, we examined several methods for choosing the probabilities used by the stochastic degree sequence model (SDSM), finding that the bipartite configuration model (BiCM) is both the fastest and most accurate. In study 2, we examined the statistical power of the SDSM relative to the fixed degree sequence model (FDSM), finding that the SDSM can be viewed as a statistically less powerful (or more conservative) variant of the FDSM. In study 3, we examined the similarity of an FDSM-extracted backbone to backbones extracted using other models, finding that the SDSM and FDSM extract very similar backbones from bipartite networks with a wide range of possible degree distributions when an appropriate significance level $\alpha$ is chosen. Finally, in study 4, we examined the ability for backbones extracted using different models to recover a known community structure, finding that although all models yield a backbone that recovers the structure, SDSM and FDSM can detect a community structure more clearly and from a weaker signal.

Based on these findings, and with the goal of offering researchers some guidance in extracting the backbones of bipartite projections, we offer three recommendations. First, \textit{we recommend the stochastic degree sequence model (SDSM) for extracting the backbones of bipartite projections} because it is fast, controls for both agent and artifact degree sequences, and yields modular backbones when the bipartite data contains even modest evidence of within-community clustering. Second, when the SDSM is used, \textit{we recommend that the cell-filling probabilities $p^*_{ik}$ be chosen using the Bipartite Configuration Model (BiCM)} because it is faster and more accurate than any other currently available method. Third, when an FDSM backbone extracted at the $\alpha = 0.05$ significance level is desired but computationally infeasible, \textit{we recommend extracting an SDSM backbone at the $\alpha = 0.13$ significance level}, which we observe is very similar when there is variation in the agent and artifact degree sequences. The models and options necessary to adopt these recommendations are implemented in the \texttt{backbone} package for \textsf{\textbf{R}}.\cite{domagalski2021backbone}

These findings and recommendations must be viewed in light of the fact that, due to the computational requirements of the FDSM and of extracting a large number of backbones across the four studies, these studies have relied on small synthetic bipartite networks ranging in size from $3 \times 3$ (study 1) to $200 \times 1000$ (study 4). However, in practice bipartite networks may be several orders of magnitude larger. For example, a bipartite network used to infer collaborations in the US House of Representatives includes 435 agents (representatives) and over 6000 artifacts (bills),\cite{neal2020,neal_domagalski_yan} while a bipartite network used to infer movie recommendations includes 17,770 agents (films) and nearly 500,000 artifacts (viewers).\cite{zweig2011systematic} Future research should explore whether these findings extend to backbones extracted from such large bipartite networks. Limitations of existing backbone models also point to directions for future research. First, using the FDSM will generally be computationally infeasible in practice because the distribution of $P^*_{ij}$ arising from $\mathcal{B}^{\text{FDSM}}$ must be estimated via numerical simulation. Identifying this distribution's probability mass function, which is known for the other ensembles (see \textit{Supplementary Text S1}), would facilitate the use of this otherwise attractive model. Second, all the ensemble models we have considered impose constraints on the degree sequences, but other types of constraints may also be useful. For example, in some contexts it may be necessary to constrain all members of an ensemble to contain a $0$ in a particular cell (e.g., to represent that an author was not alive to co-author a paper, or a legislator was not present to co-sponsor a bill).\cite{snijders1991enumeration} These limitations and future directions notwithstanding, the results presented above provide a starting point for further development of backbone models, and provide applied researchers with some practical guidance on model selection.

\bibliography{bibliography}

\section*{Acknowledgements}
This work was supported by the National Science Foundation (\#1851625 \& \#2016320) and the Michigan State University Center for Business and Social Analytics.

\section*{Author contributions statement}
Z.P.N. conceived the research questions, designed and conducted the analysis, wrote the first draft, and prepared the revisions. R.D. and Z.N. wrote the backbone package. B.S. wrote the proofs. All authors analysed the results and reviewed the manuscript.

\section*{Additional information}
\subsection*{Competing interests}
The authors declare no competing interests.

\subsection*{Supplementary information}
All code necessary to replicate these analyses is available at \url{https://osf.io/m4yfd/}. The \texttt{backbone} package used to perform the analyses is available for \textsf{R} from CRAN, and can be installed by typing \texttt{install.packages(``backbone'')} in the \textsf{R} console.

\newpage\setcounter{page}{1}
\begin{center}
\textbf{Supplementary Information for:}\\
\textbf{``Comparing Models for Extracting the Backbone of Bipartite Projections''}\\
\textbf{Zachary P. Neal, Rachel Domagalski, \& Bruce Sagan}\\
\end{center}

\section{Probability Mass Functions of projection edge weights under ensemble backbone models}
In the subsections below, we derive the probability mass functions of $P^*_{ij}$ used by ensemble backbone models to evaluate the statistical significance of the weight of edge $P_{ij}$ in a bipartite projection. We use the following notation:
\begin{itemize}
    \setlength\itemsep{-1em}
    \item Let $\mathbf{B}$ be an $m \times n$ bipartite matrix, with a vector of row sums $R = (r_1,\dots,r_m)$, a vector of column sums $C = (c_1,\dots,c_n)$, and $f$ cells containing a $1$.  So
    $$
    f = \sum_{i=1}^m r_i = \sum_{j=1}^n c_j.
    $$
    \item Let $\mathcal{B^\text{M}}$ be the ensemble of all $m \times n$ matrices $\mathbf{B^*}=(B^*_{ij})$ that obey the constraints of the respective model.  In all models, the probability distribution on $\mathcal{B^\text{M}}$ is uniform except in the stochastic case.
    \item Let $P^*_{ij}$ be a random variable equal to $(\mathbf{B^*}\mathbf{B^*}^T)_{ij}$ for all $\mathbf{B^*}~\in~\mathcal{B^\text{M}}$. Note that we have
    \begin{equation}
    \label{P*ij}
    P^*_{ij} = B^*_{i1} B^*_{j1} + B^*_{i2} B^*_{j2} + \cdots 
    + B^*_{in} B^*_{jn}.
    \end{equation}
\end{itemize}

\subsection{Fixed Fill Model (FFM)}
Let the \textit{fixed fill model} constrain all $\mathbf{B^*}~\in~\mathcal{B}^\text{FFM}$ to contain the same number of 1s (i.e. fill) as $\mathbf{B}$. 

\begin{thm}
Under the fixed fill model, the distribution of $P^*_{ij}$ for $i\neq j$ satisfies
\begin{equation}
\label{fill}
\Pr(P^*_{ij}=k) = \frac{\dil\binom{n}{k}\sum_r 2^{n-k-r} \binom{n-k}{r}\binom{(m-2)n}{f-n-k+r}}{\dil\binom{mn}{f}}.
\end{equation}
\end{thm}
\bprf
For the denominator we need to compute the cardinality 
$\# \mathcal{B}^\text{FFM}$.  If ${\bf B^*}\in \mathcal{B}^\text{FFM}$ then 
${\bf B^*}$ has $mn$ entries of which $f$ must be chosen to be ones.  So 
$$
\# \mathcal{B}^\text{FFM}=\binom{mn}{f}.
$$

For the numerator, suppose $P^*_{ij}=k$.  We see  from equation~\ree{P*ij}  that there are exactly $k$ columns $c$ where $B^*_{ic}=B^*_{jc}=1$.  There are $\binom{n}{k}$ ways to choose these columns.
Now define the following parameters:
\begin{align*}
p &= \text{number of columns $c$ where $B^*_{ic}=1$ and $B^*_{jc}=0$},\\
q &= \text{number of columns $c$ where $B^*_{ic}=0$ and $B^*_{jc}=1$},\\
r &= \text{number of columns $c$ where $B^*_{ic}=0$ and $B^*_{jc}=0$}.\\
\end{align*}
The number of ways to pick the columns counted by these parameters from the $n-k$ columns which do not contains ones in both rows is the trinomial coefficients $\binom{n-k}{p,q,r}$.  Now we have used $2k+p+q$ ones in rows $i$ and $j$.  So there are $f-2k-p-q$ left to distribute to the remaining $m-2$ rows.  And these rows have $(m-2)n$ entries.  So the number of possibilities for these remaining ones is 
$\binom{(m-2)n}{f-2k-p-q}$.  Thus the total number of choices from this and the previous paragraph is
\begin{align*}
\binom{n}{k} \sum_{p+q+r=n-k} \binom{n-k}{p,q,r}\binom{(m-2)n}{f-2k-p-q}
&=\binom{n}{k}\sum_{p+q+r=n-k}\binom{n-k}{r}\binom{n-k-r}{p}\binom{(m-2)n}{f-n-k+r}\\[5pt]
&=\binom{n}{k}\sum_r \binom{n-k}{r}\binom{(m-2)n}{f-n-k+r} \sum_p \binom{n-k-r}{p}\\[5pt]
&=\binom{n}{k}\sum_r 2^{n-k-r} \binom{n-k}{r}\binom{(m-2)n}{f-n-k+r}
\end{align*}
as desired.
\eprf

For even modestly large $\mathbf{B}$, computing equation \eqref{fill} involves values larger than can be handled by some programs. In practice, we use logs to make these computations practical.

We now show that the sum in the numerator of this probability is related to the famous Jacobi orthogonal polynomials. This sum is a terminating hypergeometric series.  Given a real number $a$ and a nonnegative integer $r$ the corresponding {\em Pochhammer symbol} or {\em rising factorial} is
$$
(a)_r = a(a+1)(a+2)\cdots(a+r-1).
$$
Note that if $a$ is an integer with $-r<a\le 0$ then $(a)_r=0$ because the product contains $0$ as a factor. Given real numbers $a_1,a_2,\ldots,a_p$ and $b_1,b_2,\ldots,b_q$ as well as a variable $z$, the corresponding {\em hypergeometric series} is
$$
_pF_q\left[
\begin{array}{cccc}
  a_1   &  a_2 & \ldots & a_p \\
  b_1   &  b_2 & \ldots & b_q
\end{array}
; z
\right]
=\sum_{r\ge0} \frac{(a_1)_r (a_2)_r \cdots (a_p)_r}
{(b_1)_r (b_2)_r \cdots (b_q)_r} \frac{z^r}{r!}.
$$
Note that if any of the $a_i$ are negative integers then, because of the remark above, this series will terminate and become a polynomial in $z$.

To convert a binomial coefficient into Pochhammer symbols, we write
\begin{align*}
\binom{n}{r}&=\frac{(n)(n-1)\cdots(n-r+1)}{r!}\\
&=\frac{(-1)^r (-n)(-n+1)\cdots(-n+r-1)}{(1)_r}\\
&=\frac{(-1)^r (-n)_r}{(1)_r}.
\end{align*}
The following identity will also be useful
\begin{align*}
(a)_{b+r} 
&= (a)(a+1)\cdots(a+b-1)\times 
(a+b)(a+b+1)\cdots (a+b+r-1)\\
&=(a)_b (a+b)_r.
\end{align*}

We now return to the sum in the numerator of equation~\eqref{fill}.  We will ignore the factor of $2^{n-k}$ since it is constant with respect to the sum and so can be pulled outside.  For simplicity of calculation we will also use the substitutions
$$
s = (m-2)n,\qquad t = f-n-k.
$$
Thus we have
\begin{align*}
\sum_r 2^{-r} \binom{n-k}{r}\binom{(m-2)n}{f-n-k+r}
&= \sum_r \binom{n-k}{r} \binom{s}{t+r} (1/2)^r\\[5pt]
&=\sum_r \frac{(-1)^r(k-n)_r}{(1)_r}
\cdot \frac{(-1)^{t+r}(-s)_{t+r}}{(1)_{t+r}} (1/2)^r\\[5pt]
&=(-1)^t\sum_r \frac{(k-n)_r (-s)_t (-s+t)_r}{(1)_t (t+1)_r} \frac{(1/2)^r}{(1)_r}\\[5pt]
&=\frac{(-1)^t (-s)_t}{(1)_t}\sum_r \frac{(k-n)_r(-s+t)_r}{(t+1)_r} \frac{(1/2)^r}{r!}\\[5pt]
&= \binom{s}{t}\  \rule{0pt}{0pt}_2F_1\left[
\begin{array}{c}
  k-n  \quad -s+t \\
  t+1
\end{array}
; \frac{1}{2}
\right]
\end{align*}

We are indebted to Marko Petkov\v{s}ek [personal communication] for pointing out  that this $_2F_1$ is, up to a factor, a specialization of a Jacobi polynomial.  Given a nonnegative integer $\ell$ and real numbers $\alpha,\beta$ the associated {\em Jacobi polynomial} is
$$
P_\ell^{(\alpha,\beta)}(z) = 
\binom{\al+\ell}{\ell}\
\rule{0pt}{0pt}_2F_1\left[
\begin{array}{c}
  -\ell \quad \ell+\alpha+\beta+1 \\
  \alpha+1
\end{array}
; \frac{1-z}{2}
\right]
$$
To make these $_2F_1$ polynomials agree we can let
$\ell=n-k$, $\alpha=t=f-n-k$, 
$$
\beta=-s+t-(\ell+\alpha+1)= k-(m-1)n-1
$$
and $z=0$.  With these substitutions we get
$$
\sum_r 2^{-r} \binom{n-k}{r}\binom{(m-2)n}{f-n-k+r}
=\frac{\dil\binom{(m-2)n}{f-n-k}}{\dil\binom{f-2k}{n-k}}\
P_{n-k}^{(f-n-k,\ k-(m-1)n-1)}(0).
$$

\subsection{Fixed Row Model (FRM)}
Let the \textit{fixed row model} constrain all $\mathbf{B^*}~\in~\mathcal{B}^\text{FRM}$ to have the same row sums as $\mathbf{B}$. 

\begin{thm}
Under the fixed row model, the distribution of $P^*_{ij}$ for $i\neq j$ is hypergeometric and satisfies 
$$\Pr(P^*_{ij}=k)=
\frac{\dil\binom{r_j}{k}\binom{n-r_j}{r_i-k}}{\dil\binom{n}{r_i}}.
$$ 
\end{thm}
\bprf
The total number of ways to pick $r_i$ of the $n$ columns for ones in the $i$th row and $r_j$ of the $n$ columns for ones in the $j$th row is 
\begin{equation}
\label{FRMbot}    
\binom{n}{r_i}\binom{n}{r_j}=\binom{n}{r_i}\frac{n!}{r_j! (n-r_j)!}.
\end{equation}
So that will go in the denominator of the desired probability.

For the numerator we follow the same line of reasoning as in the previous proof, where the parameters therein can be expressed as
\begin{align*}
p &= r_i-k,\\
q &=r_j-k,\\
r &= n-r_i-r_j+k.\\
\end{align*}
So we have a total of 
\begin{equation}
\label{FRMtop}
\binom{n}{k}\binom{n-k}{p,q,r}= \frac{n!}{k!(r_i-k)!(r_j-k)!(n-r_i-r_j+k)!}
\end{equation}
choices.

Dividing equation~\ree{FRMtop} by~\ree{FRMbot} and cancelling $n!$ gives
$$
\Pr(P^*_{ij}=k)
= \frac{\dil\frac{r_j!}{k!(r_j-k)!}\cdot\frac{(n-r_j)!}{(r_i-k)!(n-r_i-r_j+k)!}}
{\dil\binom{n}{r_i}}
=\frac{\dil\binom{r_j}{k}\binom{n-r_j}{r_i-k}}{\dil\binom{n}{r_i}}.
$$
as desired.
\eprf

\subsection{Distribution of projection edge weights under the Fixed Column Model (FCM)}
Let the \textit{fixed column model} constrain all $\mathbf{B^*}~\in~\mathcal{B}^\text{FCM}$ to have the same column sums as $\mathbf{B}$. 

Let $X_1, \ldots, X_n$ be independent Bernoulli random variables.  Let the probability of success for $X_i$ be
$$
\Pr(X_i=1) = p_i.
$$
The random variable 
\begin{equation}
\label{pbd}
X=X_1+\cdots+X_n
\end{equation}
is said to have the {\em Poisson binomial distribution} with parameters $p_1,\ldots,p_n$.

\begin{thm}
Under the fixed column model, the distribution of $P^*_{ij}$ for $i\neq j$ is Poisson binomial with parameters
$$
p_1 = \frac{c_1(c_1-1)}{m(m-1)},\
p_2 = \frac{c_2(c_2-1)}{m(m-1)},\
\ldots,\
p_n = \frac{c_n(c_n-1)}{m(m-1)}.
$$
\end{thm}
\bprf
The $B^*_{ik}$ are all either zero or one and are independent in different columns when only the column sums are fixed.  So as $k$ varies, the products $B^*_{ik}B^*_{jk}$ are independent Bernoulli random variables.  Comparing equations~\ree{P*ij} and~\ree{pbd}, we see that the distribution of $P^*_{ij}$ is Poisson binomial.

If column $k$ has column sum $c=c_k$ then all zero-one vectors with sum $c$ are equally likely for that column of ${\bf B^*}$.  So there are $\binom{m}{c}$ possible $k$th columns.  
The number of ways to have a success is the number of possible columns which have ones in both positions $i$ and $j$ where $i\neq j$.  So the number of choices is the number of ways to choose the remaining $c-2$ ones in that column from the other $m-2$  positions, that is,  $\binom{m-2}{c-2}$.  Thus
$$
p_k= \Pr(B^*_{ik}B^*_{jk}=1) 
=\frac{\dil\binom{m-2}{c-2}}{\dil\binom{m}{c}}
=\frac{c(c-1)}{m(m-1)}
$$
which finishes the demonstration.
\eprf

\subsection{Stochastic Degree Sequence Model (SDSM)}
In the \textit{stochastic degree sequence model}, $\mathcal{B}^\text{SDSM}$ consists of all binary $m\times n$ matrices.  A method is then chosen to generate probabilities $p^*_{ik}$.  Finally, matrices $\mathbf{B}^*\in\mathcal{B}^\text{SDSM}$ are generated using these probabilities for  independent Bernoulli trials, where 
$B^*_{ik}$ is filled with a one with probability $p^*_{ik}$ and zero otherwise.

\begin{thm}
Under the stochastic degree sequence model, the distribution of $P^*_{ij}$ for $i\neq j$ is Poisson binomial with parameters
$$
p_1 = p^*_{i1}p^*_{j1},\ \ldots,\ p_n=p^*_{in}p^*_{jn}.
$$
\end{thm}
\bprf
The fact that the distribution is Poisson binomial follows immediately from the independence assumption on the $\Pr(B^*_{ik})$ and equation~\ree{P*ij}.
Furthermore, the probability that the $k$th variable is one is 
$$
p_k=\Pr(B^*_{ik} B^*_{jk} = 1) = \Pr(B^*_{ik} =1) \Pr(B^*_{jk} = 1)
=p^*_{ik}p^*_{jk}.
$$
So we are done.
\eprf

\section{Familywise error rates in backbone extraction}
When testing the hypothesis that an observed statistic $s$ is different from what would be expected at random (i.e. under a given null model), the researcher must specify a significance level $\alpha$. The researcher then computes the probability $p$ of observing a value greater than or equal to $s$ under the null model. The null hypothesis is rejected and the alternative hypothesis is supported if $p < \alpha$. When only one hypothesis is being tested, this procedure ensures a Type-I error rate -- a false positive, or the risk of rejecting the null hypothesis when it is true -- of $\alpha$.

In the context of backbone extraction, the `statistic' $s$ is the number of co-occurrences between two agent nodes or the edge weight in the bipartite projection, and the `null model' is defined by the chosen bipartite ensemble backbone models. When deciding whether a given edge should be included in the backbone, the researcher is testing a single hypothesis where the null hypothesis is that the edge's weight is no stronger than would be expected under the null model. If the null hypothesis is rejected, then the edge is included in the backbone. Committing a Type-I error in this context results in including the edge in the backbone when it should be excluded (i.e. a false positive).

When multiple independent hypotheses are tested simultaneously, the Type-I error rate is inflated. Specifically, the familywise error rate $\bar{\alpha}$ -- the risk of making one or more Type-I errors -- is $1-(1-\alpha)^t$, where $t$ is the number of independent tests. For example, if the Type-I error rate for each hypothesis test is $\alpha = 0.05$, and $t = 100$ independent tests are performed, then $\bar{\alpha} = 1-(1-0.05)^{100} = 0.995$. That is, it is virtually guaranteed that at least one Type-I error will be committed in these 100 hypothesis tests. Because extracting a backbone requires the researcher to conduct a hypothesis test for every edge (with non-zero weight) in the network, backbone extraction nearly \textit{always} involves testing multiple independent hypotheses. 

Many different procedures exist for controlling $\bar{\alpha}$ when multiple independent hypothesis tests are conducted. All of these procedures involve using a corrected significance level $\alpha^*$ for each hypothesis test so that $\bar{\alpha}$ is maintained at the desired tolerance for Type-I error. The simplest but also most conservative approach is the Bonferroni correction, which defines $\alpha^* = \frac{\alpha}{t}$. Other less conservative and more powerful corrections include the Holm-Bonferroni correction \cite{holm1979simple} which has been used to extract the backbone of a political network \cite{aref2021detecting}, and the False Discovery Rate \cite{benjamini1995controlling} which has been used to extract the backbones of movie rating and international trade networks \cite{saracco2017inferring}. These correction procedures, as well as several others, are available in the \texttt{backbone} package we use to extract backbones in our studies \cite{domagalski2021backbone}. 

Using one of these procedures to control $\bar{\alpha}$ is usually appropriate when extracting the backbone of a bipartite projection. Doing so is often straightforward because (1)  many backbone models we consider (FRM, FCM, FFM, SDSM) yield exact $p$-values, and (2) the \texttt{backbone} package we use to extract backbones in our studies implements several different methods for correcting $\alpha^*$ and thus controlling $\bar{\alpha}$. However, for reasons we describe below, it is computationally infeasible to control $\bar{\alpha}$ when extracting backbones using FDSM. While this represents a significant limitation to using FDSM backbones in practice, and is a key reason we are seeking alternatives, this is not a problem for our studies. Within each of our studies, the rate of Type-I error inflation is identical for all backbones, which means that uncorrected FDSM backbones can be compared to uncorrected non-FDSM backbones. 

\subsection{Controlling FWER in FDSM backbones}
It is computationally infeasible to control $\bar{\alpha}$ when the backbone is extracted from a bipartite projection using FDSM. The challenge arises because each edge's $p$-value is estimated via a Monte Carlo procedure, and estimating these $p$-values with sufficient precision and confidence requires an impractically large number of Monte Carlo trials. In this section, we describe one way to estimate the required number of trials, and illustrate why controlling $\bar{\alpha}$ in FDSM backbones is impractical.

Because the associated probability mass function is unknown, when using FDSM to extract a backbone, the $p$-value of a given edge's weight (i.e. the probability that the same or larger edge weight would be observed under the null model) is estimated via Monte Carlo methods. Following $N$ Monte Carlo trials in which a projection $P^*$ is constructed from a random $\mathbf{B^*}~\in~\mathcal{B}^{\text{FDSM}}$, the $p$-value of the edge between $i$ and $j$ is
$$p_{ij} = \frac{\text{number of trials where } P^*_{ij} \geq P_{ij}}{N} \text{.}$$
Therefore, the estimation of $p_{ij}$ is equivalent to estimating a proportion from a sample.

Determining the sample size required to estimate a proportion from a sample with a given error tolerance is a well-studied problem in statistical inference, under the heading of power analysis. \cite{fleiss2013statistical} show that the required minimum sample $N$ to determine whether an estimated proportion $P_1$ differs from a hypothesized proportion $P_0$, with a Type-I error rate of $\epsilon_1$ and Type-II error rate of $\epsilon_2$, is
$$
N \geq\left[\frac{z_{\epsilon_1} \sqrt{P_{0}\left(1-P_{0}\right)}+z_{\epsilon_2} \sqrt{P_{1}\left(1-P_{1}\right)}}{P_{1}-P_{0}}\right]^{2}
 \text{, }
$$
where $z$ represents the critical value corresponding to $\epsilon_1$ or $\epsilon_2$ in the standard normal distribution. Note that the Type-II error rate is the opposite of the Type-I error rate, the risk of failing to rejecting the null hypothesis when it is, in fact, false (i.e. a false negative). In the backbone context, committing a Type-II error results in excluding an edge from the backbone when it should be included. They further recommend performing a minor correction to arrive at a final estimate $N'$
$$
N' \geq N + \frac{1}{|P_1 - P_0|}
\text{.}
$$

With a small adaptation to their first expression, we can use these to estimate the required number of FDSM Monte Carlo trials. We wish to use it to determine the required minimum number of Monte Carlo samples $N$ to determine whether an edge's estimated $p$-value $p_{ij}$ differs from our corrected significance level $\alpha^*$. Accordingly, we can re-write the expression given by \cite{fleiss2013statistical} as:
$$
N \geq\left[\frac{z_{\epsilon_1} \sqrt{\alpha^*\left(1-\alpha^*\right)}+z_{\epsilon_2} \sqrt{p_{ij}\left(1-p_{ij}\right)}}{p_{ij}-\alpha^*}\right]^{2}
 \text{.}
$$

Two examples serves to illustrate how this expression implies that an impractically large number of Monte Carlo trials will be required under even modest assumptions. Suppose we are using FDSM to extract the backbone from a projection of a $100 \text{ agent} \times 1000 \text{ artifact}$ bipartite network, and we wish to maintain a familywise error rate of $\bar{\alpha} = 0.05$. If we assume that our bipartite projection will be dense (hence the need for extracting its backbone) and will not contain any zero-weight edges, then we must conduct $\frac{100(100-1)}{2} = 4950$ independent hypothesis tests. Using the Bonferroni correction for simplicity of illustration, this implies a corrected two-tailed significance level of $\alpha^* = \frac{0.05/4950}{2} \approx 0.000005$ for each test. Further, assume that we are willing to tolerate a 5\% risk of incorrectly including an edge (i.e. Type-I error, $\epsilon_1 = 0.05$), and a 5\% risk of incorrectly excluding an edge (i.e. Type-II error, $\epsilon_2 = 0.05$), because both types of errors are equally problematic for graphs. 

Under these assumptions, we can consider two scenarios. First, we can determine how many (additional) trials are necessary to make a decision about an edge whose statistical significance appears unambiguous after some number of initial trials. When it appears that $p_{ij} = 0$, this represents a `best case scenario' in which it should be relatively easy to reach a decision about whether the edge should be included in the backbone. We can compute the required number of trials as:
\begin{align*}
N &\geq\left[\frac{z_{.05} \sqrt{0.000005\left(1-0.000005\right)}+z_{.05} \sqrt{0\left(1-0\right)}}{0-0.000005}\right]^{2} \\
N &\geq\left[\frac{1.64 \sqrt{0.000005\left(1-0.000005\right)}+1.64 \sqrt{0\left(1-0\right)}}{0-0.000005}\right]^{2} \\
N &\geq 535~695 \text{ (initial estimate)} \\
N' &\geq 535~695 + \frac{1}{|0-0.000005|} \\
N' &\geq 733~695 \text{ (adjusted estimate)}
\end{align*}
Under this best case scenario, at least 733,695 Monte Carlo trials are required to reach a decision given our familywise error rate and tolerances for Type-I and Type-II errors. Recall that each Monte Carlo trial requires sampling one $\mathbf{B^*}~\in~\mathcal{B}^{\text{FDSM}}$ using the curveball algorithm, then multiplying $\mathbf{B^*}$ by its transpose. Although the running time of these two operations is relatively fast (approximately 0.07 seconds on the system we use to evaluate running times in Study 1), performing the required number of trials under this best case scenario would take around 14 hours.

Second, consider a more realistic scenario in which, after some initial number of trials, an edge's statistical significance is more ambiguous because $p_{ij}$ is near $\alpha^*$. For the sake of illustration, consider an edge whose $p$-value we initially estimate as $p_{ij} = 0.0000038$, which appears smaller than $\alpha^* = 0.000005$, but is close and therefore riskier. In this case,
\begin{align*}
N &\geq 29~845~088 \text{ (initial estimate)} \\
N' &\geq 30~637~088 \text{ (adjusted estimate)}
\end{align*}
Under this more realistic scenario, where the edge's statistical significance is not unambiguous, over 30 million Monte Carlo trials are required to reach a decision given our error tolerance. This would require a running time of approximately 25 days.

\end{document}